\documentstyle[12pt]{article}

\newcommand{\be}{\begin{equation}}
\newcommand{\ee}{\end{equation}}
\newcommand{\bn}{\begin{eqnarray}}
\newcommand{\en}{\end{eqnarray}}
\newcommand{\bd}{\begin{displaymath}}
\newcommand{\ed}{\end{displaymath}}

\begin{document}

\begin{flushright}
$IFT-P.074/97$
\\hep-th/9711152
\end{flushright}
\vspace{0.5cm}
\begin{center}
{\Large \bf Path integrals on a flux cone}
\footnote{Work supported by FAPESP grant 96/12259-1.} 
\\ 
\vspace{1cm} 
{\large E. S. Moreira Jnr.}\footnote{e-mail: moreira@axp.ift.unesp.br}   
\\ 
\vspace{0.3cm} 
{\em Instituto de F\'{\i}sica  Te\'{o}rica} \\ 
{\em Universidade Estadual Paulista,}   \\
{\em Rua Pamplona, 145}  \\
{\em 01405-900 - S\~{a}o Paulo, S.P., Brazil}  \\
\vspace{0.3cm}
{\large November, 1997}
\end{center}
\vspace{1cm}

\begin{abstract}

This paper considers the Schr\"{o}dinger propagator
on a cone with the conical singularity carrying 
magnetic flux (``flux cone'').
Starting from the operator formalism and then combining techniques of path 
integration in polar coordinates and in spaces with 
constraints, the propagator and its path integral representation
are derived. ``Quantum correction" in the Lagrangian appears naturally 
and no a priori assumption is made about connectivity of the 
configuration space.

\end{abstract}

\section{Introduction}
Quantum mechanics on cones has been showing to be a
fruitful model for studying the interplay between 
quantum mechanics and geometry.
The nearly trivial geometry of the cone
(curvature is concentrated at a single point, the conical singularity
\cite{sok77,des84},
resulting that the geometry is Euclidean everywhere except on a ray
which starts at the singularity \cite{mor97})
is responsible for Aharonov-Bohm (A-B) like effects
which have been discovered throughout the years [3-7].
Such findings can be used in the study of various
(real) quantum systems whose backgrounds can be regarded as being conical 
with good approximation. Quantum matter around cosmic strings
and black holes, and
statistical mechanics of identical particles in two dimensions 
are examples.

In this paper a path integral representation for the propagator of the 
Schr\"{o}dinger equation
is derived from the operator formalism on the cone.
A magnetic flux is let to run through the cone axis,
so that one has an A-B set up coupled with the conical geometry.
The method contrasts with the one in the literature where
path integral representations in spaces with a singular point 
are obtained by 
angular decomposition of the Feynman prescription in Cartesian coordinates,
and by assuming non simply connectivity of the configuration space [8-12].
In the present approach instead, topological features arise naturally.

The paper is organized as follows.
In section 2 the background is briefly discussed
(for more detailed accounts see \cite{mor97} and references therein).
In section 3 
path integral prescription (and propagator) is derived
by breaking the evolution operator up into an infinite product of
short time evolution operators, and then inserting completeness relations
for configuration space eigenstates,
whose orthonormality relation is expressed in terms of stationary states.
(Such procedure is straightforward in Euclidean space, 
but rather elaborate in non trivial backgrounds \cite{kle90}.)
Topological features are identified
in the resulting expression.
The paper closes with final remarks.

\section{The background}

A cone is obtained from the Euclidean plane
by removing a wedge of angle 
$2\pi \left(1-\alpha \right)$ 
(in fact when $\alpha>1$ a wedge is inserted).
Clearly the line element is given by 
\be
dl^{2}=d\rho^{2}+\rho^{2}d\varphi^{2}, 
\label{dlP}
\ee
which is the line element of the Euclidean plane written in polar coordinates.
The fact that there is a delta function curvature at the origin is encoded
in the unusual identification
\be
\left(\rho ,\varphi \right) \sim  \left(\rho ,\varphi +2\pi\alpha \right). 
\label{Boun}
\ee

The behavior of a free  particle with mass $M$ 
on a cone is determined from the Lagrangian, 
\bn
{\cal L}&=&\frac{1}{2}M  \left(dl/dt \right)^{2}
\nonumber
\\
&=&\frac{1}{2} M \left( \dot{ \rho}^{2} + 
\rho ^{2} \dot{\varphi }^{2} \right).
\label{lagg}
\en
Noting (\ref{Boun}) it follows that
orbits of particles (geodesic motion on the cone) 
are simply broken straight lines with uniform motion.
As a constant magnetic flux $\Phi$ running through the cone axis
does not affect classical motion of a particle (with charge $e$), 
then classical motion on a flux cone is nearly trivial.
Quantum motion, on the other hand, reveals
non trivial features \cite{aha59}. 

%he following sections consider the quantum
%motion interpreted in the context of the path integral approach.
%It should be mention that path integrals aspects of the 
%Schr\"{o}dinger propagators on cones
%have also been considered in \cite{dow77,ger90}.

\section{The propagator and its path integral representation}
Due to local flatness of the conical geometry 
the free Hamiltonian operator is just the free Hamiltonian
operator on the plane,
\be
H= -\frac{\hbar^{2}}{2M}\frac{1}{\rho}\frac{\partial}{\partial
\rho}\left(\rho\frac{\partial}{\partial \rho}\right) +\frac{ L^{2}}
{2M\rho^{2}},
\label{hamo}
\ee
where $L:=-i\hbar \partial/\partial \varphi$.
By choosing an appropriate gauge
(the one corresponding to a vector potential which vanishes
everywhere, except on a ray)
and observing (\ref{Boun}),
it follows that solutions of the 
Schr\"{o}dinger equation satisfy \cite{mor97}
\be
\psi(\rho,\varphi+2\pi\alpha)=\exp\{i2\pi\sigma\}\psi(\rho,
\varphi),
\label{boun}
\ee
with $\sigma:= -e\Phi/ch$.
Boundary condition (\ref{boun}) carries all information
about the non trivial geometry and magnetic field.

Consider the following effective Lagrangian 
\be
{\cal L}_{eff}=
\frac{M}{2}\left(\dot{\rho}^{2} + \rho ^{2} \dot{\varphi}^{2}\right)+
\frac{\hbar^{2}}{8M\rho ^{2}},
\label{effl}
\ee
which is obtained from (\ref{lagg}) by adding a 
quantum correction.
The corresponding Hamiltonian is given by
\be
{\cal H}_{eff}= \frac{1}{2M}\left (p_{\rho}^{2}
+\frac{p_{\varphi}^{2}}{\rho ^{2}}-\frac{\hbar ^{2}}
{4\rho ^{2}} \right). 
\label{effh}
\ee
The momentum operators associated with $p_{\rho}$
and with $p_{\varphi}$ are given by
\bn
p_{\rho}\rightarrow -i\hbar 
\left(\partial _{\rho}+ \frac{1}{2\rho}\right)
&&p_{\varphi}\rightarrow L,
\label{srv}
\en
where the presence of the term $-i\hbar/2\rho$
ensures self-adjointness of $p_{\rho}$ (if the wave functions do not
diverge very rapidly at $\rho=0$ \cite{mor97}), without spoiling the usual 
canonical commutation relations \cite{mar80,kle90}. 
It turns out that by performing the substitutions (\ref{srv})
in (\ref{effh}), the Hamiltonian operator (\ref{hamo}) is reproduced,
which obviously would not be the case if the quantum correction
was not present in (\ref{effh}) \cite{art69}. 
The effective Lagrangian (\ref{effl}) will be considered
again below.

One seeks stationary states which span a space of wave 
functions where conservation of probability holds.
This implies that the singularity 
at the origin must not be a source or a sink,
\be
\lim_{\rho \rightarrow 0}
\int_{0}^{2\pi\alpha}d\varphi\
\rho {\rm J}_{\rho}=0,
\label{pc}
\ee
where ${\rm J}_{\rho}$ is the usual expression
for the radial component of the probability current
on the plane.
Condition (\ref{pc}) is automatically guaranteed if the 
stationary states are finite at the origin.
(Mildly divergent boundary conditions can be equally 
compatible with conservation of probability and 
square integrability of the wave function
\cite{kay91,bou92,mor97}. These possibilities will not be considered here.)
Functions
\bn
\psi_{k,m}(\rho,\varphi)&=&\langle\rho,\varphi|k,m\rangle\nonumber\\
&=& \frac{1}{\sqrt{2\pi\alpha}}J_{|m+\xi|/\alpha}(k\rho)
e^{i(m+\xi)\varphi/\alpha},
\label{eig}
\en
where 
$0\leq k < \infty$,  $m$ 
is a integer and $J_{\nu}$ denotes a Bessel function of the first kind,
are simultaneous eigenfunctions of
$H $ and $ L$ 
with eigenvalues 
$\hbar^{2} k^{2}/2M$ and $(m+\xi)\hbar/\alpha$ 
respectively. Note that since $J_{\nu}(0)$ is finite for non negative $\nu$,
these stationary states are finite at the origin.

States $|\rho,\varphi\rangle$  
in (\ref{eig}) are a complete set of configuration space
eigenstates,
\be
\int_{0}^{\infty}d\rho\ \rho \int_{0}^{2\pi\alpha}d\varphi\ 
|\rho,\varphi\rangle\langle\rho,\varphi| = {\bf 1}.
\label{cr}
\ee
Since their orthonormality relation is 
\bd
\langle\rho,\varphi|\rho ' ,\varphi '\rangle =\frac{1}{\rho}
\delta(\rho-\rho ')\delta(\varphi-\varphi '),
\ed
it follows that
\be
\langle\rho,\varphi|\rho ' ,\varphi '\rangle =
\sum_{m=-\infty}^{\infty}\int_{0}^{\infty}dk\ k
\psi_{k,m}(\rho ,\varphi) 
\psi^{\ast}_{k,m}(\rho ',\varphi ').
\label{crst}
\ee
This expression is the completeness relation of the eigenfunctions
$\psi_{k,m}$,
%\linebreak[4]
\bd
\sum_{m=-\infty}^{\infty}\int_{0}^{\infty}dk\ k|k,m\rangle\langle k,m|
={\bf 1}, 
\ed
which may be derived by using
the completeness relation of the Bessel functions,
\be
\int_{0}^{\infty}dk\ kJ_{\nu}(k\rho)J_{\nu}(k\rho')=
\frac{1}{\rho}\delta(\rho-\rho'),
\label{crb}
\ee
with Poisson's formula,
\be
\sum_{m=-\infty}^{\infty}\delta(\phi +2\pi m)=
\frac{1}{2\pi}\sum_{m=-\infty}^{\infty}\exp\left\{im\phi\right\}.
\label{pf}
\ee
Expression (\ref{crst})
corresponds to the usual one in  Cartesian coordinates
where  
$\langle {x}|{x'}\rangle$ is expressed
in terms of plane waves, 
$\langle {x}|{x'}\rangle=
\int \left(dk/2\pi\right)
\exp\left\{ik\left(x-x'\right)\right\}$.
Recall that plane waves are
simultaneous eigenfunctions 
of the free Hamiltonian  and linear momentum operators,
whereas $\psi_{k,m}(\rho ,\varphi)$ are
simultaneous eigenfunctions of the free Hamiltonian and
angular momentum operators.

The orthonormality relation for the eigenfunctions $\psi_{k,m}$,
\bn
\langle k,m|k',m'\rangle&=&
\int_{0}^{\infty}d\rho\ \rho\int_{0}^{2\pi\alpha}
d\varphi\ 
\psi^{\ast}_{k,m}(\rho ,\varphi) 
\psi_{k',m'}(\rho,\varphi)
\nonumber
\\&=&
\frac{1}{k}\delta(k-k')\delta _{mm'},
\label{orst}
\en
follows from the orthonormality relation
\be
\int_{0}^{2\pi\alpha}d\varphi\ \exp\{i\varphi (m-n)/\alpha\}=
2\pi\alpha\delta _{mn} 
\label{orrr}
\ee
and (\ref{crb}).

For a complete set of configuration space eigenstates
$|\rho,\varphi\rangle$ 
the propagator of the Schr\"{o}dinger equation is given by      
$K \left(\rho,\varphi;\rho ',\varphi' ;\tau \right) =
\langle\rho,\varphi|U\left( \tau \right) |\rho',\varphi'\rangle$, where
$ U(\tau):=\exp\{-iH\tau/\hbar\}$ 
is the evolution operator and 
$\tau := t-t'$ is the time interval.
Slicing $\tau$ in $N+1$ slices of width
$\epsilon=\tau_{n}=t_{n}-t_{n-1}=\tau/(N+1)$, 
the propagator reads
\be
K \left(\rho,\varphi;\rho ',\varphi' ;\tau \right) =
\langle\rho,\varphi|\prod_{n=1}^{N+1} U\left( \tau_{n} \right) 
|\rho',\varphi'\rangle,
\label{pro}
\ee
where the composition law of the evolution operator was used
with the identifications 
$t\equiv t_{N+1}$ and $t'\equiv t_{0}$.
By inserting in (\ref{pro}) $N$ completeness relations (\ref{cr})
between each pair of evolution operators, one is led to
\be
K \left(\rho,\varphi;\rho ',\varphi' ;\tau \right) =
\prod_{n=1}^{N}\left[\int_{0}^{\infty}d\rho_{n}\ 
\rho_{n}\int_{0}^{2\pi\alpha}
d\varphi_{n}\right]
\prod_{n=1}^{N+1}\left[\langle\rho_{n},\varphi_{n}| U(\epsilon)
|\rho_{n-1},\varphi_{n-1}\rangle\right].
\label{Pro}
\ee
The identifications
$|\rho,\varphi\rangle\equiv|\rho_{N+1},\varphi_{N+1}\rangle$
and 
$|\rho',\varphi'\rangle\equiv|\rho_{0},\varphi_{0}\rangle$
were also used.

The short time amplitudes in (\ref{Pro}) may be rewritten as
\be
\langle\rho_{n},\varphi_{n}|U(\epsilon)|\rho_{n-1},\varphi_{n-1}\rangle=
\langle\rho_{n},\varphi_{n}|\rho_{n-1},\varphi_{n-1}\rangle
-i\frac{\epsilon}{\hbar}H
\langle\rho_{n},\varphi_{n}|\rho_{n-1},\varphi_{n-1}\rangle
+{\cal O}(\epsilon^{2}).
\label{samp}
\ee
In order to obtain the action of $H$ on 
$\langle\rho_{n},\varphi_{n}|\rho_{n-1},\varphi_{n-1}\rangle$
one expresses the later in terms of eigenfunctions of the former, i.e.
(\ref{crst}) is considered. Then (\ref{samp}) is recast as
\bn
\langle\rho_{n},\varphi_{n}|U(\epsilon)|\rho_{n-1},\varphi_{n-1}\rangle
&=&
\sum_{m=-\infty}^{\infty}\int_{0}^{\infty}dk\ k
e^{-i\epsilon E_{k}/\hbar}
\psi_{k,m}(\rho_{n},\varphi_{n}) 
\psi^{\ast}_{k,m}(\rho_{n-1},\varphi_{n-1}) \nonumber
%\label{}
\\
&&+\quad {\cal O}(\epsilon^{2}),
\label{Samp}
\en
where $E_{k}$ denotes the eigenvalue of $H$, i.e. $\hbar^{2}k^{2}/2M$.
% in (\ref{eigenvalues}).
By replacing (\ref{Samp}) in (\ref{Pro}) and taking the limit 
$N\rightarrow  \infty$ ($\epsilon \rightarrow 0$)
a partitioned expression for the propagator is obtained,
\bn
K \left(\rho,\varphi;\rho ',\varphi' ;\tau \right)&=&
\lim_{N\rightarrow \infty}
\prod_{n=1}^{N}\left[\int_{0}^
{\infty}d\rho_{n}\ \rho_{n}\int_{0}^{2\pi\alpha}
d\varphi_{n}\right]\label{pro3a}\\
&&\times\prod_{n=1}^{N+1}
\left[\sum_{m=-\infty}^{\infty}\int_{0}^{\infty}dk\ k
e^{-i\epsilon E_{k}/\hbar}
\psi_{k,m}(\rho_{n},\varphi_{n}) 
\psi^{\ast}_{k,m}(\rho_{n-1},\varphi_{n-1})\right].
%\label{}
\nonumber
\en
The integral over $k$ in (\ref{pro3a}) may be evaluated by 
using the formula \cite{grad80}
\be
\int_{0}^{\infty}dx\ xe^{-ax^{2}}
J_{\nu}(bx)J_{\nu}(cx)=
(1/2a)e^{-(b^{2}+c^{2})/4a}I_{\nu}(bc/2a), 
\label{int}
\ee
where ${\rm Re}\ a > 0$, ${\rm Re}\ \nu > -1$.
This integral corresponds in Cartesian coordinates to the Gaussian integral.
Analytic continuation of (\ref{int}) gives 
\bn
&&K \left(\rho,\varphi;\rho ',\varphi' ;\tau \right)=
\lim_{N\rightarrow \infty}
\frac{M}{2\pi\alpha i\epsilon\hbar}
\prod_{n=1}^{N}\left[\int_{0}^
{\infty}d\rho_{n}\ \rho_{n}\int_{0}^{2\pi\alpha}
\frac{d\varphi_{n}}{2\pi\alpha i\epsilon\hbar/M}\right]
\label{pro4a}\\
&&\quad\qquad\qquad\times\prod_{n=1}^{N+1}
\left[e^{iM\left(\rho_{n}^{2}+\rho_{n-1}^{2}\right)/2\hbar\epsilon}
\sum_{m=-\infty}^{\infty}I_{|m+\sigma|/\alpha}
\left(M\rho_{n}\rho_{n-1}/i\hbar\epsilon\right)
e^{i(m+\sigma)\left(\varphi_{n}-\varphi_{n-1}\right)/\alpha}
\right]
%\label{}
\nonumber
\en

When $\sigma$ is an integer and the space is Euclidean, i.e. $\alpha=1$, 
(\ref{pro4a}) reduces to Feynman's prescription for the propagator 
of a free particle. Indeed by considering the Fourier expansion of a plane
wave,
\be
\exp\left\{ia\cos\phi\right\}=
\sum_{m=-\infty}^{\infty}I_{|m|}(ia)e^{im\phi},
\label{rpw}
\ee
one sees from (\ref{pro4a}) the familiar partitioned expression 
\be
K_{0} \left({\bf x},{\bf x}';\tau \right)=
\lim_{N\rightarrow \infty}
\frac{M}{2\pi i\epsilon\hbar}
\prod_{n=1}^{N}\left[\int
\frac{d^{2}x_{n}}{2\pi i\epsilon\hbar/M}\right]
\exp\left\{\frac{i}{\hbar}\sum_{n=1}^{N+1}\epsilon\frac{M}{2}
\left(\frac{{\bf x}_{n}-{\bf x}_{n-1}}{\epsilon}\right)
^{2}\right\},
\label{pro5}
\ee
which is symbolically written as
\be
K_{0}\left({\bf x},{\bf x}';\tau \right)=
\int {\cal D}^{2}x\ \exp\left\{\frac{i}{\hbar}\int_{t'}^{t}
dt\ \frac{M}{2}\dot{{\bf x}}^{2}\right\}.
\label{sym1}
\ee

Before rewriting the path integral representation 
(\ref{pro4a}) in a symbolic form which is analogous
to (\ref{sym1}), the
expression for the propagator on the flux cone 
which was obtained in \cite{dow77} using
a complex contour method will be reproduced here
from (\ref{pro3a}). (References \cite{des88,ger90} have
also reproduced this propagator when $\sigma=0$ using other
methods. Reference \cite{ger90} in particular has used a
path integral approach which is a generalization to the cone
of the method used in the A-B set up \cite{ber81}.
The propagator when $\alpha=1$ has been long known in the
literature \cite{kre65}.) 
Observing (\ref{orst}), it is seen that
only one sum over $m$ and one integration over $k$ remain in
(\ref{pro3a}),
\bn
K\left(\rho,\varphi;\rho ',\varphi' ;\tau \right)&=&
\frac{1}{2\pi\alpha}
\label{rpro3a}
\int_{0}^{\infty}dk\ k
e^{-i\tau E_{k}/\hbar}\\
&&\times\sum_{m=-\infty}^{\infty}
\nonumber
J_{|m+\sigma|/\alpha}\left(k\rho\right)
J_{|m+\sigma|/\alpha}\left(k\rho '\right)
e^{i(m+\sigma)\left(\varphi-\varphi'\right)/\alpha}.
\en
Then using (\ref{int}) to  evaluate
the integration over $k$, results in
\be
K \left(\rho,\varphi;\rho ',\varphi' ;\tau \right)=
\frac{M}{2\pi\alpha i\tau\hbar}
\label{pro6a}
e^{iM\left(\rho^{2}+\rho'^{2}\right)/2\hbar\tau}
\sum_{m=-\infty}^{\infty}I_{|m+\sigma|/\alpha}
\left(M\rho\rho'/i\hbar\tau\right)
e^{i(m+\sigma)\left(\varphi-\varphi'\right)/\alpha},
\ee
which could have been guessed from (\ref{pro4a}).
From (\ref{rpw}) it follows that
when $\sigma$ is an integer and $\alpha=1$, 
(\ref{pro6a}) collapses into the free Schr\"{o}dinger
propagator on the Euclidean plane, viz.
\bd
K_{0} \left({\bf x},{\bf x}';\tau \right)=
\frac{M}{2\pi i\tau\hbar}
e^{iM\left({\bf x}-{\bf x}'\right)^{2}/2\hbar\tau}.
\ed

Noting that 
$\int_{-\infty}^{\infty}d\lambda\  I_{\lambda}(z)
\delta (\lambda -\nu) \exp\{i\lambda\phi\}=
I_{\nu}(z)\exp\{i\nu\phi\}$
and using (\ref{pf}), (\ref{pro6a}) becomes
\be
K \left(\rho,\varphi;\rho ',\varphi' ;\tau \right)=
\sum_{l=-\infty}^{\infty}
e^{-i2\pi l\sigma} 
\tilde{K}\left(\rho,\varphi+2\pi\alpha l ;\rho ',\varphi' ;\tau \right),
\label{pro7}
\ee
with 
\be
\tilde{K}\left(\rho,\varphi;\rho ',\varphi' ;\tau \right):=
\frac{M}{2\pi i\tau\hbar}
e^{iM\left(\rho^{2}+\rho'^{2}\right)/2\hbar\tau}
\int_{-\infty}^{\infty}d\lambda\ I_{|\lambda|}
\left(M\rho\rho'/i\hbar\tau\right)
e^{i\lambda\left(\varphi-\varphi '\right)}.
\label{pro8}
\ee
Likewise (\ref{pro4a}) may be rewritten as
\bn
&&K \left(\rho,\varphi;\rho ',\varphi' ;\tau \right)=
\lim_{N\rightarrow \infty}
\frac{M}{2\pi i\epsilon\hbar}
\prod_{n=1}^{N}\left[\int_{0}^
{\infty}d\rho_{n}\ \rho_{n}\int_{0}^{2\pi\alpha}
\frac{d\varphi_{n}}{2\pi i\epsilon\hbar/M}\right]
\label{pro9}
\\&&\quad\quad\times\prod_{n=1}^{N+1}
\left[\sum_{l=-\infty}^{\infty}e^{-i2\pi l\sigma}
e^{iM\left(\rho_{n}^{2}+\rho_{n-1}^{2}\right)/2\hbar\epsilon}
\int_{-\infty}^{\infty}d\lambda\ I_{|\lambda|}
\left(M\rho_{n}\rho_{n-1}/i\hbar\epsilon\right)
e^{i\lambda\left(\varphi_{n}-\varphi_{n-1}+2\pi\alpha l\right)}\right].
\nonumber
\en
Now, by using
\bn
&&\sum_{k,l=-\infty }^{\infty }e^{(k+l)z}
\int_{0}^{c}dx\ f(kc+x)g(lc-x)=
\nonumber\\
&&\qquad\qquad\qquad\qquad\qquad\qquad\qquad
\sum_{l=-\infty}^{\infty}e^{lz}
\int_{-\infty}^{\infty}dx\ f(x)g(lc-x),
\nonumber
\en
one may extend the range of integration of $\varphi $ 
from $[0, 2\pi \alpha)$ to $(-\infty, \infty )$. 
This leaves only one sum in (\ref{pro9}),
leading to (\ref{pro7}), but now 
$\tilde{K}\left(\rho,\varphi+2\pi\alpha l;\rho ',\varphi' ;\tau \right)$
is given as a partitioned expression, 
\bn
&&\tilde{K}\left(\rho,\varphi+2\pi\alpha l;\rho ',\varphi' ;\tau \right)=
\lim_{N\rightarrow \infty}
\frac{M}{2\pi i\epsilon\hbar}
\prod_{n=1}^{N}\left[\int_{0}^
{\infty}d\rho_{n}\ \rho_{n}\int_{-\infty}^{\infty}
\frac{d\varphi_{n}}{2\pi i\epsilon\hbar/M}\right]
\label{pro10}
\\&&\quad\quad\quad\quad\times\prod_{n=1}^{N+1}
\left[e^{iM\left(\rho_{n}^{2}+\rho_{n-1}^{2}\right)/2\hbar\epsilon}
\int_{-\infty}^{\infty}d\lambda\ I_{|\lambda|}
\left(M\rho_{n}\rho_{n-1}/i\hbar\epsilon\right)
e^{i\lambda\left(\varphi_{n}-\varphi_{n-1}+
2\pi\alpha l\delta _{n, N+1}\right)}\right].
\nonumber
\en
Now the asymptotic behaviour of $I_{\nu}(z)$
for large $|z|$ can be used to derive \cite{ber81}
\bd
\int_{-\infty}^{\infty}d\lambda\ I_{|\lambda |}(z)
\exp\{i\lambda \phi\}\approx
\exp\{z+1/8z-z\phi ^{2}/2\},
\ed 
which when used in (\ref{pro10}) finally gives 
\bn
&&\tilde{K}\left(\rho,\varphi+2\pi\alpha l;\rho ',\varphi' ;\tau \right)=
\lim_{N\rightarrow \infty}
\nonumber
\frac{M}{2\pi i\epsilon\hbar}
\prod_{n=1}^{N}\left[\int_{0}^
{\infty}d\rho_{n}\ \rho_{n}\int_{-\infty}^{\infty}
\frac{d\varphi_{n}}{2\pi i\epsilon\hbar/M}\right]
\\&&
\nonumber
\exp\left\{
\frac{i}{\hbar}\sum_{n=1}^{N+1}
\epsilon
\left[\frac{M}{2}\left(\left(\frac{\rho_{n}-\rho_{n-1}}
{\epsilon}\right)^{2}+\rho_{n}\rho_{n-1}
\left(\frac{\varphi_{n}+2\pi\alpha l\delta_{n,N+1}-
\varphi _{n-1}}{\epsilon}\right)^{2}\right)\right.\right.
\\&&
\qquad\qquad\qquad\qquad\qquad\qquad +\left.\left. 
\frac{\hbar ^{2}}{8M\rho_{n}\rho_{n-1}}\right]\right\},
\label{pro11}
\en
or symbolically
\be
\tilde{K}\left(\rho,\varphi+2\pi\alpha l;\rho ',\varphi' ;\tau \right)=
\int_{0}^{\infty}{\cal D}\rho\ \rho\int_{-\infty}
^{\infty}{\cal D}\varphi\ \exp\left\{\frac{i}{\hbar}
\int_{t'}^{t}dt\left[\frac{M}{2}
\left(\dot{\rho}^{2} + \rho ^{2} \dot{\varphi}^{2}\right)+
\frac{\hbar^{2}}{8M\rho ^{2}}\right]\right\}.
\label{sym2}
\ee

\section{Final remarks}
Expressions (\ref{pro7}) and (\ref{sym2}) are 
the path integral prescription where
the corresponding action is the one made up of the effective
Lagrangian ${\cal L}_{eff}$, (\ref{effl}). Recall that 
${\cal L}_{eff}$ is the appropriate Lagrangian for quantization
through the ``substitution principle'' (\ref{srv}).
It is important to note that a na\"{\i}ve 
change from Cartesian to polar coordinates in
the Feynman prescription (\ref{sym1}) does not lead to (\ref{sym2}),
since the ``quantum correction''
$\hbar^{2}/8M\rho ^{2}$
would be missing. (Quantum corrections as this one are typical
of path integrals in non trivial backgrounds \cite{mar80}.)
This is a simple
example showing that coordinate transformations within path integral
representations raise subtle issues.

Examining expressions (\ref{pro7}) and (\ref{sym2})
leads to the following interpretation of this path
integral representation.
Since there is a conical singularity and/or
a magnetic flux at the origin,
the configuration space is non simply connected.
The propagator is given
by a sum of modulated propagators, each one of them
giving the contribution of all paths belonging
to a homotopy class labeled by
the winding number $l$.
Then the sum over $l$ in (\ref{pro7}) takes into 
account all paths circling round the ``hole'' at the origin.
The modulated factors are a unitary representation of the 
fundamental group $Z$, and the particle travels in the covering
space of $R^{2}-\{ 0\}$. The particle is not free, but interacts
with the ``non trivial'' topology through the 
quantum correction in the effective Lagrangian ${\cal L}_{eff}$.

Recalling a study of quantum flow in \cite{mor97}, one 
sees that this interpretation may be appropriate when
$\alpha<1$ and/or  $\sigma$ is a non integer. But, strictly
speaking, it is incorrect when $\alpha\geq 1$ and $\sigma$ is an integer.
In particular, when $\alpha=1$ and $\sigma=0$, (\ref{pro7})
and (\ref{sym2}) are just a polar coordinate path integral
prescription for a free particle moving on the Euclidean plane
- the apparent non trivial topology is imparted by the use of 
polar coordinates which are singular at the origin.

In principle, the material in this paper
may be reconsidered in the context of other possible boundary conditions
at the singularity. The result of such an investigation might reveal 
different features from the ones seen here. Proceeding as in section 3,
the crucial point would be the use of new stationary states 
to obtain the new propagators and their corresponding path integral
representations. This procedure seems to answer a question in 
\cite{kay91}, namely, how different boundary conditions at the singularity
are related to the path integral approach. The use of the present method
in the context of other geometries is also worth investigating.

Using the proper time representation for the Green functions, 
the extension of the method to second quantization is straightforward. 
It would be interesting to investigate the connections between 
this paper and reference \cite{ort97} where path integrals
in black hole background are considered.

\vspace{5 mm}
{\bf Acknowledgements}.
The author is grateful to George Matsas for reviewing
the manuscript.

\end{document}